\documentclass[aps,pre,groupedaddress,preprint]{revtex4}
\usepackage{graphicx}
\usepackage[dvips]{epsfig}
\usepackage{dcolumn}
\usepackage{subfigure}%
\usepackage{bm}
\usepackage{amssymb}
\usepackage{amsmath}

\begin{document}
\title{Some open questions concerning biological growth}

\author{Carlos Escudero}

\affiliation{ICMAT (CSIC-UAM-UC3M-UCM), Departamento de
Matem\'{a}ticas, Facultad de Ciencias, Universidad Aut\'{o}noma de
Madrid, Ciudad Universitaria de Cantoblanco, 28049 Madrid, Spain}

\begin{abstract}
We briefly review the properties of radially growing interfaces
and their connection to biological growth. We focus on simplified
models which result from the abstraction of only considering
domain growth and not the interface curvature. Linear equations
can be exactly solved and the phenomenology of growth can be
inferred from the explicit solutions. Nonlinear equations pose
interesting open questions that are summarized herein.
\end{abstract}

\maketitle

\section{Introduction}

Since the origins of the interdisciplinary exchange among
mathematics and biology, the development of the form and structure
of living organisms has always been considered a fascinating
topic~\cite{thompson,turing}. Already present in these seminal
works, the necessity of considering physical and mathematical
laws, together with the relevant biological principles, has been
increasingly recognized in the scientific literature along the
years. One such example is the examination of the properties
relating to the architecture of cell colonies, as already noted by
Murray Eden~\cite{eden1,eden2}. The overall appearance of a living
organism is conditioned by its genetic expression. Indeed, it
escapes to nobody that twins are similar to each other as whole
individuals. However, when we regard smaller scale properties of
them, like the dermal ridge count of fingerprints or the patterns
of retinal venation, their differences are greater even in the
case of monozygotic twins. Differences that are expected to grow
as our reference spatial scale decreases, and that are
particularly evident in the architecture of cell colonies. These
and other observations unveil that morphogenetic processes are not
completely determined by the genetic expression. Of course, it is
widely recognized nowadays that environmental factors play a
determinant role in morphogenesis. Both aside and as a consequence
of this one would be interested in determining how random factors
affect the process of growth, and so how is this process
constraint by the laws of probability.

The Eden model was introduced in order to shed some light into
this question~\cite{eden1,eden2}. It is a simplified model for
biological growth, which results from the abstraction of
neglecting many of the real aspects of cell colonies development.
It concentrates just in the appearance of new cells in the colony
periphery; once introduced, cells are never removed from it. This
model is to be considered in $(\mathbb{Z}^d)^{\{0,1\}}$, where $d
\in \mathbb{N}$ denotes the spatial dimension, and $0$ and $1$
stand for an empty and an occupied site respectively. Starting
from a single cell at the origin, and following a set of
probabilistic rules which dictates the frequency and manner in
which new cells are introduced in the colony~\cite{eden1,eden2}, a
radial form develops in the long time. Its macroscopic shape is
affected by the underlying lattice structure, and its interface
shows fractal properties that are independent of
it~\cite{barabasi,halpin}.

Although a rigorous connection have never been established, it is
widely accepted that the fluctuations of the Eden interface can be
described with some suitable stochastic partial differential
equation~\cite{barabasi,halpin,hammersley}. Such a relation has
been proposed in light of numerical results derived with the so
called cylindrical Eden model~\cite{hammersley}. In this case, for
$d=2$, the system is a strip of infinite length and finite width
$L$, and the initial condition is a whole semistrip. For large
enough $L$ and periodic boundary conditions the system obeys the
Family-Vicsek ansatz~\cite{family}, i. e., the two-points
correlation function is of the form
\begin{equation}
\langle h(x,t) h(x',t) \rangle = |x-x'|^{2 \alpha} f \left(
\frac{|x-x'|}{t^{1/z}} \right),
\end{equation}
where $\langle \cdot \rangle$ is the average over a large number
of realizations, $f(\cdot)$ is the scaling function, and $\alpha$
and $z$ are universal quantities known as critical exponents:
$\alpha$ is known as the roughness exponent and $z$ as the
dynamical exponent, which measures the velocity at which the
correlations travel. The function $h(x,t)$ expresses the height of
the interface with respect to the initial condition at some given
position $x$ and time $t$. The ratio of the critical exponents
$\beta=\alpha/z$ constitutes a new exponent measuring the velocity
at which the interface width increases during the first stages of
growth~\cite{barabasi,halpin}. For the dimensionality considered
the simulations have measured $\alpha \approx 1/2$ and $z \approx
3/2$, which place this discrete model in the same universality
class as the continuum equation
\begin{equation}
\partial_t h= \nu \nabla^2 h + \frac{\lambda}{2} (\nabla h)^2 + F + \xi(x,t),
\end{equation}
which is known as the Kardar-Parisi-Zhang (KPZ)
equation~\cite{kpz}. In this case $\xi(x,t)$ is a Gaussian
distributed spatiotemporal noise delta correlated in both space
and time, and $\nu$, $\lambda$ and $F$ are real positive
parameters. Of course, the cylindrical Eden model differs from its
radial counterpart in two features: the interface is curved and
grows laterally. Herein we will consider an abstraction of this
problem and focus exclusively on lateral growth. As discrete
models are usually placed in the universality classes defined by
continuum equations, we will study the dynamics of such equations
defined on uniformly growing domains.

\section{Linear growth}

In this section we will summarize some of the recent results
obtained with linear equations. As they are exactly solvable their
dynamical structure can be inferred from explicit expressions.

Our study of the dynamics of stochastic growth equations on
growing domains begins with a stochastically forced diffusion
equation, known as the Edwards-Wilkinson (EW) equation~\cite{EW}
in this context, which reads
\begin{equation}
\partial_t h=D\nabla^2 h + F + \xi({\bf y},t),
\end{equation}
where $\xi({\bf y},t)$ is a zero-mean Gaussian white noise which
correlation is
\begin{equation}
\langle \xi({\bf y},t)\xi({\bf y}',t') \rangle=\epsilon
\delta({\bf y}-{\bf y}')\delta(t-t'),
\end{equation}
$D$ is the diffusion constant, $F$ the constant rate at which mass
enters the interface and $\epsilon$ the noise intensity, all these
parameters being positive real numbers. We start considering the
conservation law in integral form
\begin{equation}
\frac{d}{dt}\int_{S_t} h({\bf y},t)d{\bf y}=\int_{S_t}\left[
-\nabla \cdot {\bf j}+ \mathcal{F}({\bf y},t) \right]d{\bf y},
\end{equation}
where $S_t$ is the uniformly growing domain, ${\bf j}=-D\nabla h$
is the current generated by diffusion, and $\mathcal{F}({\bf
y},t)=F+\xi({\bf y},t)$ is the EW growth
mechanism~\cite{escudero1}. By applying the Reynolds transport
theorem we find
\begin{equation}
\frac{d}{dt}\int_{S_t}h({\bf y},t)d{\bf
y}=\int_{S_t}\left[\partial_t h + \nabla \cdot ({\bf v}h) \right]
d{\bf y},
\end{equation}
where ${\bf v}({\bf y},t)$ denotes the flow velocity generated by
the growing domain. Valid as it is for any domain, the integral
conservation law may be expressed in the local form
\begin{equation}
\label{local}
\partial_t h+\nabla \cdot ({\bf v}h)=D\nabla^2h+\mathcal{F}({\bf y},t).
\end{equation}
In this equation we readily identify two new terms, the advection
one ${\bf v}\cdot \nabla h$, and the dilution one $h \nabla \cdot
{\bf v}$. For every ${\bf y} \in S_t$, that has evolved from ${\bf
y}_0 \in S_{t_0}$, we find ${\bf v}({\bf y},t)=\partial {\bf
y}/\partial t$. Let us now concentrate in one-dimensional domains
and then move to higher dimensions. In this case uniform growth
translates into $y=g(t)y_0$, where $g(t)$ is a temporal function
such that $g(t_0)=1$. This yields $v=y\dot{g}/g$, and thus
\begin{equation}
\partial_t h +\frac{\dot{g}}{g}\left( y\partial_y h + h \right)= D \partial_y^2 h + F + \xi(y,t).
\end{equation}
For a one-dimensional domain $\left(0,L(t)\right)$, with
$L(t)=g(t)L_0$, we change the spatial coordinate $x = y L_0/L(t)$,
where $L_0=L(t_0)$, in order to map the problem into the interval
$\left( 0,L_0 \right)$. This transformation counterbalances
advection, and so the resulting equation reads
\begin{equation}
\frac{\partial h}{\partial t}=\left(\frac{L_0}{L(t)}\right)^2D
\frac{\partial^2 h}{\partial x^2}-\frac{\dot{g}}{g}h+F+
\sqrt{\frac{L_0}{L(t)}} \xi(x,t),
\end{equation}
where we have used the fact that the noise is delta correlated.
The dilution term has become $h \nabla \cdot {\bf v}=
-(\dot{g}/g)h$. Dilution has a transparent physical meaning: as
the domain grows the incoming mass becomes distributed in a larger
($d-$dimensional) area. Now we assume that the growth function
adopts the power law form $g(t)=(t/t_0)^\gamma$, where the growth
index $\gamma \ge 0$, to find
\begin{equation}
\label{gamma} \frac{\partial h}{\partial
t}=\left(\frac{t_0}{t}\right)^{2\gamma} D \frac{\partial^2
h}{\partial x^2}-\frac{\gamma}{t}h+F+
\left(\frac{t_0}{t}\right)^{\gamma/2} \xi(x,t).
\end{equation}
The growth index $\gamma$ is a new degree of freedom of this
problem; it cannot be deduced from the other model parameters, and
has to be measured directly from the physical system under study.

Now we move to a more general situation in which we consider an
arbitrary diffusion operator of order $\zeta$ and an arbitrary
spatial dimension $d$, see Eq. (\ref{local}). From now on the
$d-$dimensional coordinates will be denoted ${\bf x} \to x$ and
${\bf y} \to y$ for simplicity. In this case, we can proceed
exactly in the same way as in the one-dimensional situation to
find, instead of Eq. (\ref{gamma}), the equation
\begin{equation}
\label{gammazeta}
\partial_t h= -D \left( \frac{t_0}{t} \right)^{\zeta \gamma} |\nabla|^\zeta h
-\frac{d\gamma}{t}h+F+\left(\frac{t_0}{t}\right)^{d\gamma/2}\xi(x,t),
\end{equation}
where the fractional operator $|\nabla|^\zeta$ is to be understood
in terms of the Fourier transform. Special values of $\zeta$ yield
some of the well known equations in this topic, as the EW equation
for $\zeta=2$~\cite{EW} and the Mullins-Herring equation for
$\zeta=4$~\cite{mh1,mh2}. There is still another way of deriving
an equation similar to (\ref{gammazeta}) but in which the dilution
term is not present. If we just considered the dilatation
transformation $x \to (t/t_0)^\gamma x$ instead of domain growth
we would find
\begin{equation}
\label{gammazeta2}
\partial_t h= -D \left( \frac{t_0}{t} \right)^{\zeta \gamma} |\nabla|^\zeta h
+F+\left(\frac{t_0}{t}\right)^{d\gamma/2}\xi(x,t).
\end{equation}
The difference among equations (\ref{gammazeta}) and
(\ref{gammazeta2}) appears already in the amount of mass arriving
at the interface~\cite{escudero2}. In absence of external sources
of mass, i. e. $F=\epsilon=0$, and for no flux boundary conditions
the total mass on the surface is conserved
\begin{equation}
\int_0^{L(t)} \cdots \int_0^{L(t)}h(y,t)dy = \int_0^{L_0} \cdots
\int_0^{L_0}h(x,t_0)dx,
\end{equation}
for the dilution dynamics (\ref{gammazeta}). In the no dilution
situation corresponding to equation (\ref{gammazeta2}) we find
\begin{equation}
\label{zeromode}
\int_0^{L(t)} \cdots \int_0^{L(t)}h(y,t)dy =
\left( \frac{t}{t_0} \right)^{d \gamma} \int_0^{L_0} \cdots
\int_0^{L_0}h(x,t_0)dx.
\end{equation}
This second case, as we have already mentioned, is pure
dilatation, which implies that not only the space grows, but also
the interfacial matter grows at the same rate, in such a way that
the average density remains constant. Note that this process of
matter dilatation, as well as the spatial growth, are
deterministic processes.

We have analyzed both types of dynamics, in the absence and
presence of dilution, and found a number of measurable
consequences. Temporal dynamics can be studied by means of the
temporal auto-correlation
\begin{equation}
\label{temporalc} A(t,t') \equiv \frac{\langle
h(x,t)h(x,t')\rangle_0}{\langle h(x,t)^2 \rangle_0^{1/2} \langle
h(x,t')^2 \rangle_0^{1/2}} \sim
\left(\frac{\min\{t,t'\}}{\max\{t,t'\}}\right)^\lambda, \quad
\mathrm{for} \quad \max\{t,t'\} \gg \min\{t,t'\},
\end{equation}
where $\lambda$ is the auto-correlation exponent and $\langle
\cdot \rangle_0$ denotes the average with the zeroth mode
contribution suppressed. When dilution is considered the
auto-correlation exponent takes the form
\begin{equation}
\lambda = \left\{ \begin{array}{ll} \beta + d/\zeta &
\mbox{\qquad if \qquad $\gamma < 1/\zeta$}, \\
\beta +\gamma d & \mbox{\qquad if \qquad $\gamma
> 1/\zeta$}, \end{array} \right.
\end{equation}
or alternatively
\begin{equation}
\lambda= \beta + {d \over z_\lambda},
\end{equation}
where $\beta= 1/2-d/(2\zeta)$ and
\begin{equation}
z_\lambda = \min\{\zeta,1/\gamma\}.
\end{equation}
In absence of dilution we find
\begin{equation}
\lambda = \left\{ \begin{array}{ll} {1 \over 2} +\frac{d}{2\zeta}
-d
\gamma & \mbox{\qquad if \qquad $\gamma < 1/\zeta$}, \\
{1 \over 2}-\frac{d}{2\zeta} & \mbox{\qquad if \qquad $\gamma
> 1/\zeta$}, \end{array} \right.
\end{equation}
or alternatively
\begin{equation}
\lambda= \beta + {d \over z_{\mathrm{eff}}},
\end{equation}
where $\beta= 1/2-d/(2\zeta)$ and
\begin{equation}
z_{\mathrm{eff}} = \left\{ \begin{array}{ll} \zeta/(1-\gamma
\zeta) & \mbox{\qquad if \qquad $\gamma < 1/\zeta$}, \\
\infty & \mbox{\qquad if \qquad $\gamma
> 1/\zeta$}. \end{array} \right.
\end{equation}
From these formulas one can clearly read that when dilution is
suppressed there is no mechanism for correlations propagation.
Otherwise dilution is the responsible for the propagation of
correlations in the fast growth regime, i. e. when $\gamma >
1/\zeta$.

Complementary information can be obtained from the interface
persistence. The persistence of a stochastic process denotes its
tendency to continue in its current state. When considering the
dynamics of a fluctuating interface, one refers to the persistence
probability $P_{+}(t_1,t_2)$ ($P_{-}(t_1,t_2)$) as the pointwise
probability that the interface remains above (below) its profile
at $t_1$ up to time $t_2>t_1$ \cite{kallabis,krug}. Herein, as in
\cite{singha}, we concentrate on the case in which the initial
profile is flat, and we suppress the contribution coming from the
zeroth mode again. For the stochastic differential equations under
consideration the symmetry $h_n \to -h_n$ for all Fourier modes $n
\neq 0$ holds, implying the equality $P_+ = P_- \equiv P$. For
long times $t_2 \gg t_1$ we have the power law behavior
\cite{kallabis,krug}
\begin{equation}
P(t_1,t_2) \sim (t_1/t_2)^\theta,
\end{equation}
defining the persistence exponent $\theta$. It was previously
calculated in the limit $\zeta \to \infty$ when $\gamma=0$
\cite{krug}
\begin{equation}
\theta \approx \frac{1}{2}+\frac{2\sqrt{2}-1}{2}\frac{d}{\zeta},
\end{equation}
up to higher order terms, and in this same limit when $d=1$ and
$\gamma=1$ \cite{singha}
\begin{equation}
\theta \approx \frac{1}{2}-\frac{1}{2\zeta},
\end{equation}
up to higher order terms and in the absence of dilution. We have
calculated the persistence exponent~\cite{escudero1}, again in the
limit $\zeta \to \infty$ and assuming the inequality $\gamma >
1/\zeta$, in the presence of dilution
\begin{equation}
\theta \approx \frac{1}{2} +d\gamma - \frac{d}{2\zeta},
\end{equation}
up to higher order terms, and in the absence of it
\begin{equation}
\theta \approx {1 \over 2} -\frac{d}{2 \zeta},
\end{equation}
generalizing the previous result \cite{singha}. The exponent
$\theta=1/2$ characterizes neutrally persistent interfaces, which
are those deprived of a relaxation mechanism (i. e. $D=0$ in
equation (\ref{gammazeta2})). For $\theta < 1/2$ the interface is
persistent and for $\theta > 1/2$ it is antipersistent. Note that
if dilution acts on the interface then it is antipersistent, as in
the case of no domain growth; contrarily, if dilution is not
present, the interface becomes persistent.

Before we start calculating spatial correlations let us note that
domain growth induces the length scale $|x-x'| \sim t^{(1-\zeta
\gamma)/\zeta}$. First we show the scaling form that the two
points correlation function adopts for ``microscopic'' spatial
scales $|x-x'| \ll t^{(1-\zeta \gamma)/\zeta}$ in the fast growth
regime. In this case one has~\cite{escudero2}
\begin{equation}
\langle h(x,t)h(x',t) \rangle - \langle h(x,t) \rangle^2  \approx
|x-x'|^{\zeta-d} t^{\gamma(\zeta-d)}\mathcal{F}\left[ |x-x'|
t^{(\zeta \gamma -1)/\zeta} \right],
\end{equation}
or in Lagrangian coordinates $|y-y'|=|x-x'|t^\gamma$
\begin{equation}
\label{scaling} \langle h(y,t)h(y',t) \rangle - \langle h(y,t)
\rangle^2 \approx |y-y'|^{\zeta-d} \mathcal{F}\left[
\frac{|y-y'|}{t^{1/\zeta}} \right],
\end{equation}
where we have assumed the inequality $\zeta > d$ and the
statistical isotropy and homogeneity of the system in the limit in
which the scaling form holds. As dilution does not act on such a
microscopic scale, these results are independent of whether we
contemplate dilution or not. Things are different for macroscopic
length scales $|x-x'| \gg t^{(1-\zeta \gamma)/\zeta}$. In this
limit dilution has a measurable action, and when it is included in
the interface equation of motion the resulting correlation
is~\cite{escudero1}
\begin{equation}
\langle h(y,t)h(y',t) \rangle - \langle h(y,t) \rangle^2  \sim t
\delta(y-y'),
\end{equation}
which is simply the short time limit of equation (\ref{scaling}).
If dilution is suppressed we find
however~\cite{escudero4,escudero5}
\begin{equation}
\label{aroughness}
\langle h(y,t)h(y',t) \rangle_0 \sim \left\{
\begin{array}{lll} t \, \delta(y-y') & \mbox{\qquad if \qquad $\gamma <
1/d$}, \\ t \, \ln(t) \, \delta(y-y') & \mbox{\qquad if \qquad $\gamma = 1/d$}, \\
t^{\gamma d} \, \delta(y-y') & \mbox{\qquad if \qquad $\gamma >
1/d$}.
\end{array} \right.
\end{equation}
In this case we see that for fast enough growth memory effects
appear and modify the time dependent
prefactor~\cite{escudero4,escudero5}. The increase of this
prefactor reflects the mass excess that enters the interface when
dilution is not operating as shown in equation (\ref{zeromode}). A
consequence of all these correlations is the scale dependent
fractal dimension
\begin{equation}
d_f(|x-x'|,t) = \left\{ \begin{array}{ll} 1+(3d-\zeta)/2 &
\mbox{\qquad if \qquad $|x-x'| \ll t^{(1-\zeta \gamma)/\zeta}$},
\\
d+1 & \mbox{\qquad if \qquad $|x-x'| \gg t^{(1-\zeta
\gamma)/\zeta}$}, \end{array} \right.
\end{equation}
which is independent of whether we contemplate dilution or not,
and reveals the interface multifractality. These asymptotic values
suggest the self-similar form of the fractal dimension
\begin{equation}
d_f=d_f \left( \frac{|x-x'|}{t^{(1-\zeta \gamma)/\zeta}} \right),
\end{equation}
which would imply its invariance with respect to the dilatation $x
\to b \, x$, $t \to b^{z_f}t$, and $d_f \to b^{\alpha_f} d_f$, for
$z_f=\zeta/(1-\zeta \gamma)$, $\alpha_f=0$ and $b$ a real number
strictly greater than one.

As a final note let us mention that the assumption $\zeta > d$ is
fundamental in order to get the correlations specified by equation
(\ref{scaling}). As we have seen, in this case the dynamical
exponent is universal and given by $z=\zeta$. For $\gamma=0$ and
$\zeta=d=1$ the dynamical exponent is still universal and given by
$z=1$; however for $\gamma=\zeta=d=1$ and in the absence of
dilution this exponent becomes non-universal and given by
$z=F/(D+F) \in (0,1)$~\cite{escudero3}. The presence of dilution
restores universality and the non-growing domain result
$z=1$~\cite{escudero2}.

\section{Nonlinear growth}

The open questions in this topic are related, not surprisingly, to
the appearance of nonlinear terms in the corresponding equations
of motion. One of the most popular nonlinear models in this
context is the KPZ equation, which as we have commented in the
Introduction is related to the biologically motivated Eden model.
As we will see, understanding the KPZ equation on a growing domain
may shed some light on some of the properties of the classical
version of this model.

The KPZ equation on a growing domain reads~\cite{escudero2}
\begin{equation}
\label{dkpz}
\partial_t h= \nu \left( \frac{t_0}{t} \right)^{2 \gamma} \nabla^2 h +
\frac{\lambda}{2} \left( \frac{t_0}{t} \right)^{2 \gamma} (\nabla
h)^2 -\frac{d \gamma}{t}h + \gamma Ft^{\gamma-1} + \left(
\frac{t_0}{t} \right)^{d \gamma/2} \xi(x,t).
\end{equation}
Of course, if we just considered the dilatation $x \to
(t/t_0)^\gamma x$ we would find
\begin{equation}
\label{ndkpz}
\partial_t h= \nu \left( \frac{t_0}{t} \right)^{2 \gamma} \nabla^2 h +
\frac{\lambda}{2} \left( \frac{t_0}{t} \right)^{2 \gamma} (\nabla
h)^2 + \gamma Ft^{\gamma-1} + \left( \frac{t_0}{t} \right)^{d
\gamma/2} \xi(x,t).
\end{equation}
As we have shown in the previous section, the dilution mechanism
fixes the Family-Vicsek ansatz in the fast growth regime. In the
radial Eden model case, assuming it belongs to the KPZ
universality class, we would have $z=3/2$ in $d=1$ and $\gamma=1$.
And so, one would na\"{\i}fly expect that the resulting interface
is uncorrelated and we have to resort on dilution effects in order
to fix the Family-Vicsek ansatz and get rid of memory effects. But
here comes the paradoxical situation. There are two main
symmetries associated with the $d$-dimensional KPZ equation: the
Hopf-Cole transformation which maps it onto the noisy diffusion
equation~\cite{wio} and Galilean invariance which have been
traditionally related to the non-renormalization of the KPZ vertex
at an arbitrary order in the perturbation expansion~\cite{fns}. In
the case of the no-dilution KPZ equation (\ref{ndkpz}) both
symmetries are still present. Indeed, this equation transforms
under the Hopf-Cole transformation $u=\exp[\lambda h/(2\nu)]$ to
\begin{equation}
\partial_t u = \nu \left( \frac{t_0}{t} \right)^{2 \gamma} \nabla^2 u +
\frac{\gamma F \lambda}{2 \nu} t^{\gamma-1} u + \frac{\lambda}{2
\nu} \left( \frac{t_0}{t} \right)^{d \gamma/2} \xi(x,t) u,
\end{equation}
which is again a noisy diffusion equation and it can be explicitly
solved in the deterministic limit $\epsilon=0$. We find in this
case
\begin{equation}
u(x,t)=\frac{(1-2\gamma)^{d/2} \exp[F \lambda t^\gamma/(2
\nu)]}{[4 \pi t_0^{2\gamma}(t^{1-2\gamma}-t_0^{1-2\gamma})]^{d/2}}
\int_{\mathbb{R}^d} \exp \left[ -\frac{|x-y|^2(1-2\gamma)}{4
t_0^{2 \gamma}(t^{1-2\gamma}-t_0^{1-2\gamma})} \right] u(y,t_0)
dy,
\end{equation}
which corresponds to
\begin{equation}
h(x,t)= \frac{2 \nu}{\lambda} \ln \left \{ \frac{(1-2\gamma)^{d/2}
\exp[F \lambda t^\gamma/(2 \nu)]}{[4 \pi
t_0^{2\gamma}(t^{1-2\gamma}-t_0^{1-2\gamma})]^{d/2}}
\int_{\mathbb{R}^d} \exp \left[ -\frac{|x-y|^2(1-2\gamma)}{4
t_0^{2 \gamma}(t^{1-2\gamma}-t_0^{1-2\gamma})} +\frac{\lambda}{2
\nu} h(y,t_0) \right] dy \right\},
\end{equation}
for given initial conditions $u(x,t_0)$ and $h(x,t_0)$. It is
clear by regarding this formula that decorrelation at the
deterministic level will happen for $\gamma > 1/2$. It is still
necessary to find out if at the stochastic level this threshold
will be moved to $\gamma > 2/3$. If we consider the dilution KPZ
equation (\ref{dkpz}) then transforming Hopf-Cole we would find
the nonlinear equation
\begin{equation}
\partial_t u = \nu \left( \frac{t_0}{t} \right)^{2 \gamma} \nabla^2
u -\frac{d \gamma}{t} u \ln(u) + \frac{\gamma F \lambda}{2 \nu}
t^{\gamma-1} u + \frac{\lambda}{2 \nu} \left( \frac{t_0}{t}
\right)^{d \gamma/2} \xi(x,t) u,
\end{equation}
which may be thought of as a time dependent and spatially
distributed version of the Gompertz differential
equation~\cite{gompertz}. In this case it is not evident how to
find an explicit solution at the deterministic level and what
would be its decorrelation threshold.

Galilean invariance means that the transformation
\begin{equation}
x \to x-\lambda v t, \qquad h \to h+vx, \qquad F \to F -
\frac{\lambda}{2}v^2,
\end{equation}
where $v$ is an arbitrary constant vector field, leaves the KPZ
equation invariant. In case of no dilution this transformation can
be replaced by
\begin{equation}
x \to x-\frac{\lambda}{1-2\gamma} v t_0^{2 \gamma} t^{1-2\gamma},
\qquad h \to h+vx, \qquad F \to F - \frac{\lambda}{2 \gamma}v^2
t_0^{2 \gamma} t^{1-3\gamma},
\end{equation}
which leaves invariant equation~(\ref{ndkpz}). If we consider
dilution, then it is not clear how to extend this transformation
to leave equation~(\ref{dkpz}) invariant. The main difficulty
comes from the dilution term which yields a non-homogeneous
contribution to the dynamics as a response to the transformation
$h \to h+vx$. So in summary we may talk of a certain sort of
Galilean invariance which is obeyed by the no-dilution KPZ
dynamics (\ref{ndkpz}) and is lost when dilution is taken into
account. If it were found that the dilution equation~(\ref{dkpz})
obeys the traditional KPZ scaling (at least in some suitable
limit), then that would put into question the role that Galilean
invariance has in fixing the exponents. The KPZ critical exponents
are believed to obey the scaling relation $\alpha + z=2$ in all
spatial dimensions, a relation that has been traditionally
attributed to Galilean invariance, although this interpretation
has been recently put into
question~\cite{hochberg1,hochberg2,wio1,wio2}.

There is still another fundamental symmetry of the KPZ equation,
but this time it just manifests itself in one spatial dimension:
the so called fluctuation-dissipation
theorem~\cite{barabasi,halpin}. It basically says that for long
times, when saturation has already being achieved, the
nonlinearity ceases to be operative and the resulting interface
profile would be statistically indistinguishable from that created
by the EW equation. For fast domain growth, we know from the
linear theory that the interface never becomes correlated, and it
operates, in this sense, as if it were effectively in the short
time regime for all times~\cite{escudero1}. As a consequence, the
fluctuation-dissipation theorem is not expected to play any role
in this case. Of course, this result would be independent of
whether we contemplated dilution or not.

\section{Disorder}

It is quite natural to consider propagation into disordered media
in the context of biological growth. One could, for instance,
imagine the development of a bacterial infection inside a host
body. This is, of course, the propagation of a bacterial front
inside a medium with a extremely low degree of symmetry.

Apart from the classical KPZ equation which is driven by thermal
noise, a different version in which this noise is replaced by
quenched disorder has been considered in the
literature~\cite{barabasi,halpin}
\begin{eqnarray}
\label{kpzpin}
&\partial_t h= \nu \nabla^2 h + \frac{\lambda}{2} (\nabla h)^2 + F + \xi(x,h),& \\
&\left< \xi(x,h) \right> = 0,& \\
&\left< \xi(x,h) \xi(x',h') \right> = \epsilon \delta(x-x')
\delta(h-h'),&
\end{eqnarray}
which is nonlinear even when $\lambda=0$. All the problems
considered in this section are open in this simpler case as well.
A way of understanding this equation is considering the simplified
random deposition version of it
\begin{equation}
\partial_t h= F + \xi(x,h),
\end{equation}
which is actually an ordinary differential equation in which the
position $x$ acts just as a parameter, at least if the quenched
disorder is conveniently regularized. And thus, let us consider
the auxiliary problem
\begin{eqnarray}
\frac{dh}{dt} &=& F + \eta(h), \\
\left< \eta(h) \right> &=& 0, \\
\left< \eta(h) \eta(h') \right> &=& \epsilon
\delta(h-h').
\end{eqnarray}
It is similar to the stochastic
problem
\begin{eqnarray}
\frac{dh}{dt} &=& F + \eta(t), \\
\left< \eta(t) \right> &=& 0, \\
\left< \eta(t) \eta(t') \right> &=& \epsilon
\delta(t-t'),
\end{eqnarray}
which solution is
\begin{equation}
h(t) = h(t_0) + F (t-t_0) + \sqrt{\epsilon} \,\, W(t-t_0),
\end{equation}
where $t_0$ is the initial time and $W(t)$ is a Wiener process, so
we basically have two superposed motions: constant drift and
Brownian motion. Classical Brownian motion describes a particle
choosing its direction of motion randomly every time step. If the
noise is position dependent, i. e. $\eta = \eta(h)$, then the
direction of motion is already prescribed in every spatial point.
For a discrete version of this process, say an unbiased random
walk on $\mathbb{Z}$ for which the jump direction is specified in
every site, if the system revisits any location, then it is
trapped forever (jumping forth and back in the last two visited
sites). The only way to prevent (this of other sort of) trapping
is to consider a sufficiently large $F$, so the system is evolving
over new positions all of the time. Something similar happens in
equation~(\ref{kpzpin}): for values of $F$ smaller than a critical
one $F_c$ the interface becomes pinned, while for larger values
the interface propagates~\cite{barabasi}.

It would be interesting to analyze the interplay of spatial
disorder with a growing domain size. We again have two
possibilities, the equation with dilution
\begin{equation}
\partial_t h= \nu \left( \frac{t_0}{t} \right)^{2 \gamma} \nabla^2 h +
\frac{\lambda}{2} \left( \frac{t_0}{t} \right)^{2 \gamma} (\nabla
h)^2 -\frac{d \gamma}{t}h + \gamma Ft^{\gamma-1} + \left(
\frac{t_0}{t} \right)^{d \gamma/2} \xi(x,h),
\end{equation}
and the one without it
\begin{equation}
\partial_t h= \nu \left( \frac{t_0}{t} \right)^{2 \gamma} \nabla^2 h +
\frac{\lambda}{2} \left( \frac{t_0}{t} \right)^{2 \gamma} (\nabla
h)^2 + \gamma Ft^{\gamma-1} + \left( \frac{t_0}{t} \right)^{d
\gamma/2} \xi(x,h).
\end{equation}
Note that, contrary to the classical case in which the interface
propagates linearly in time, in this case the velocity of
propagation would be $\sim t^\gamma$. It would be interesting to
clarify whether $\gamma=1$ plays a critical role in the dynamics
or not. In other words, whether the interface is always pinned for
$\gamma < 1$ and always moving for $\gamma>1$, or if there is a
dependence on the parameters values in these cases too. If the
second situation held, it would be yet necessary to clarify
whether there are other possible critical values for $\gamma$.
According to the microscopic description commented in this
section, it seems plausible that $\gamma=1$ is indeed the critical
value of the growth index separating pinned and unpinned regimes.

Even when $\gamma=1$ there is an interesting question associated
with dilution. We know that dilution keeps constant the amount of
matter on the interface, while suppressing it we get a mass
excess. For linearly in time growing interfaces we know that the
value of $F$, which describes the amount of matter arriving at the
interface, controls the possibility of interface
pinning/unpinning. If no dilution is present, the mass excess
could act as effectively increasing the value of $F$, and thus
facilitating interface unpinning. It would be interesting to
quantitatively determine how much the threshold of pinning is
moved in the absence of dilution, if this is indeed the case.

\section{Summary and Conclusions}

As we have seen, a consequence of the linear theory is that
dilution erases the memory effects and this way restores the
classical Family-Vicsek ansatz~\cite{escudero4,escudero5}.
Otherwise, for fast domain growth, a series of unexpected
consequences arise, as the modification of the random deposition
correlation, the lost of the antipersistent character of the
fluctuating interface and even the appearance of non-universal
critical exponents. In this respect, dilution can be thought of as
the mechanism which maintains some of the most characteristic
features of surface growth when we let the domain size grow in
time.

On the other hand, some of the well known symmetries of the KPZ
equation, as Galilean invariance and mapping to the directed
polymer problem, are maintained in a dilating setting but lost by
virtue of dilution. So, in principle, one would expect that in the
absence of dilution memory effects could be present in the KPZ
dynamics, and this way some its characteristic features would be
lost. However, it is somehow paradoxical that it is exactly this
absence of dilution what maintains the classical symmetries of
this equation. Furthermore, these symmetries have been sometimes
considered as necessary ingredients in the resulting KPZ scaling.
If we found that the KPZ equation in a rapidly growing domain and
in presence of dilution behaved in a similar way to its classical
counterpart, that would suggest that the symmetries present in the
standard situation are not playing such a necessary role.

As we have already mentioned, the motivation for studying radial
growth models such as the Eden or different ones partially comes
from the possible similarity of these with some forms of
biological development, such as for instance bacterial colonies
formation. The results of our study can be translated into this
context to obtain some simple conclusions, provided the modelling
assumptions make sense for some biological system. The structure
of a rapidly developing bacterial colony would be dominated by
dilution effects, originated in the birth of new cells which
volume causes the displacement of the existent cells. If the rate
of growth is large enough this motion will dominate over any
possible random dispersal of the bacteria. It is remarkable that
such a consequence simply appears by considering domain growth,
while it is not necessary to introduce corrections coming from the
finite size of the constituents. This is the dilution dominated
situation we have formalized by means of the (decorrelation)
inequality $\gamma > 1/\zeta$. If we were to introduce some
control protocol in order to keep the consequences of bacterial
propagation to a minimum we would need to eliminate colony
constituents (possibly randomly selected) at a high enough rate so
the effective growth velocity were one that reversed the
decorrelation inequality. For the two dimensional radial Eden
model, accepting it belongs to the KPZ universality class, one
finds $\gamma=1$ and $z =3/2$. If $z$ played the same role for the
nonlinear KPZ equation as $\zeta$ for the linear equations
considered herein (as it is reasonable to expect), the Eden model
would be in the dilution dominated regime. In order to control it
we would need to eliminate its cells at rate such that the
effective growth rate obeyed $\gamma < 2/3$. For the three
dimensional Eden model, if its behavior were still analogous to
that of the KPZ equation, we would find $z > 3/2$ and thus a
greater difficulty for control. Note that for the particular
growth rules of the Eden model one would need to eliminate
peripheral cells in order to control the system. This would not be
so in the case of an actual bacterial colony, for which bulk cells
are still able to reproduce, and so cell elimination could be
performed randomly across the whole colony. Of course, these
conclusions are speculative as long as stochastic growth equations
are not proved to reasonably model some biological system. With
respect to the problem of the experimental verification of the
theoretical results, one would be interested in finding a method
for measuring the exponents. These could perhaps be measured using
the long time dependence of the interface variance on the initial
system size and time
\begin{equation}
\left< h(x,t)^2 \right> - \left< h(x,t) \right>^2 \sim \left\{
\begin{array}{ll}
L_0^{2 \alpha} \,\, t^{2 \beta} \sim L(t)^{2 \alpha} \,\, t^{2
\beta -2
\gamma \alpha} \qquad \mathrm{for} \qquad \gamma > 1/z \\
L_0^{2 \alpha} t^{2 \gamma \alpha} \sim L(t)^{2 \alpha} \qquad
\mathrm{for} \qquad \gamma < 1/z
\end{array} \right.
\end{equation}
that we found for both the dilution and no-dilution linear
dynamics in~\cite{escudero1}. This way one could in principle
experimentally determine both exponents $\alpha$ and $\beta$ and
as a consequence the ratio $z = \alpha/\beta$ in the fast growth
regime. Subtracting the inverse of the so obtained value for the
dynamic exponent $z$ from the measured value of the growth index
$\gamma$, which should be easily obtainable from experiments, one
could estimate the distance from correlation, and in turn the
possible necessary strength of a control protocol.

\section*{Acknowledgments}

This work has been partially supported by the MICINN (Spain)
through Project No. MTM2008-03754.


\begin{thebibliography} {99}

\bibitem{thompson} D. W. Thompson, {\it On Growth and Form} (Cambridge University Press, Cambridge, 1917).

\bibitem{turing} A. M. Turing, Phil. Trans. R. Soc. B {\bf 237}, 37 (1952).

\bibitem{eden1} M. Eden, in {\it Symposium on Information Theory in Biology}, edited by H. P. Yockey (Pergamon Press, New York, 1958).

\bibitem{eden2} M. Eden, in {\it Proceedings of the Fourth Berkeley Symposium on Mathematical Statistics and Probability},
edited by J. Neyman (University of California Press, Berkeley,
1961).

\bibitem{barabasi} A.-L. Barab\'asi and H. E. Stanley, {\it Fractal Concepts in Surface Growth} (Cambridge University Press, Cambridge, 1995).

\bibitem{halpin} T. Halpin-Healy and Y.-C. Zhang, Phys. Rep. {\bf
254}, 215 (1995).

\bibitem{hammersley} J. M. Hammersley and G. Mazzarino, Comb. Probab. Comput. {\bf
3}, 471 (1994).

\bibitem{family} F. Family and T. Vicsek, J. Phys. A {\bf 18}, L75
(1985).

\bibitem{kpz} M. Kardar, G. Parisi, and Y.-C. Zhang, Phys. Rev.
Lett. {\bf 56}, 889 (1986).

\bibitem{EW} S. F. Edwards and D. R. Wilkinson, Proc. R. Soc. London
Ser. A {\bf 381}, 17 (1982).

\bibitem{escudero1} C. Escudero, J. Stat. Mech. P07020 (2009).

\bibitem{mh1} W. W. Mullins, J. Appl. Phys. {\bf 28}, 333 (1957).

\bibitem{mh2} C. Herring, J. Appl. Phys. {\bf 21}, 301 (1950).

\bibitem{escudero2} C. Escudero, arXiv:0909.5304.

\bibitem{singha} S. B. Singha, J. Stat. Mech. P08006 (2005).

\bibitem{kallabis} H. Kallabis and J. Krug, Europhys. Lett. {\bf
45}, 20 (1999).

\bibitem{krug} J. Krug, H. Kallabis, S. N. Majumdar, S. J. Cornell,
A. J. Bray, and C. Sire, Phys. Rev. E {\bf 56}, 2702 (1997).

\bibitem{singha} S. B. Singha, J. Stat. Mech. P08006 (2005).

\bibitem{escudero3} C. Escudero, Ann. Phys. {\bf 324}, 1796 (2009).

\bibitem{escudero4} C. Escudero, Phys. Rev. Lett. {\bf 100}, 116101 (2008).

\bibitem{escudero5} C. Escudero, arXiv:0907.0898.

\bibitem{wio} H. S. Wio, Int. J. Bif. Chaos {\bf 19}, 2813
(2009).

\bibitem{fns} D. Forster, D. R. Nelson, and M. J. Stephen, Phys. Rev. A {\bf 16}, 732 (1977).

\bibitem{gompertz} D. S. Jones and B. D. Sleeman, {\it Differential Equations and
Mathematical Biology}, (CRC Press, London, 2003).

\bibitem{hochberg1} A. Berera and D. Hochberg, Phys. Rev. Lett. {\bf 99}, 254501 (2007).

\bibitem{hochberg2} A. Berera and D. Hochberg, Nucl. Phys. B {\bf 814}, 522 (2009).

\bibitem{wio1} H. S. Wio, J. A. Revelli, R. R. Deza, and C. Escudero, preprint.

\bibitem{wio2} H. S. Wio, J. A. Revelli, R. R. Deza, C. Escudero, and M. S. de La
Lama, preprint.

\end{thebibliography}
\end{document}